# RADIATION-HARD ASICS FOR OPTICAL DATA TRANSMISSION IN THE ATLAS PIXEL DETECTOR

K.K. GAN, K.E. ARMS, M. JOHNSON, H. KAGAN, R. KASS, A. RAHIMI,
C. RUSH, S. SMITH, R. TER-ANTONIAN, M.M. ZOELLER

*Department of Physics, The Ohio State University, Columbus, OH 43210, USA*

A. CILIOX, M. HOLDER, S. NDERITU, M. ZIOLKOWSKI

*Fachbereich Physik, Universitaet Siegen, 57068 Siegen, Germany*

We have developed two radiation-hard ASICs for optical data transmission in the ATLAS pixel detector at the LHC at CERN: a driver chip for a Vertical Cavity Surface Emitting Laser (VCSEL) diode for 80 Mbit/s data transmission from the detector, and a Bi-Phase Mark decoder chip to recover the control data and 40 MHz clock received optically by a PIN diode. We have successfully implemented both ASICs in 0.25 µm CMOS technology using enclosed layout transistors and guard rings for increased radiation hardness. We present results from circuits of final design and from irradiation studies with 24 GeV protons up to 80 Mrad ($2.6 \times 10^{15}$ p/cm$^2$).

## 1. Introduction

The ATLAS pixel detector [1] consists of two barrel layers and two forward and backward disks which provide at least two space point measurements. The pixel sensors are read out by front-end electronics controlled by the Module Control Chip (MCC). The low voltage differential signal (LVDS) from the MCC chip is converted by the VCSEL Driver Chip (VDC) into a single-ended signal appropriate to drive a VCSEL. The optical signal from the VCSEL is transmitted to the Readout Device (ROD) via a fiber.

The ROD transmits via a fiber to a PIN diode a 40 MHz beam crossing clock bi-phase mark (BPM) encoded with the data (command) signal to control the pixel detector. This BPM encoded signal is decoded using a Digital Opto-Receiver Integrated Circuit (DORIC). The clock and data signals recovered by the DORIC are in LVDS form for interfacing with the MCC chip.

The ATLAS pixel optical readout system contains 448 VDC and 360 DORIC chips with each chip having four channels. The chips will be mounted on 180 carrier boards (opto-boards). We use beryllium oxide (BeO) as the board substrate for heat management. The optical link circuitry will be exposed to a maximum total fluence of $10^{15}$ 1-MeV $n_{eq}$/cm$^2$ during ten years of operation at the LHC. In this paper we describe the development of the radiation-hard VDC and DORIC circuits.

## 2. VDC Circuit

The VDC is used to convert an LVDS input signal into a single-ended signal appropriate to drive a VCSEL in a common cathode array. The output current of the VDC is to be variable between 0 and 20 mA through an external control current, with a standing current (dim current) of ~1 mA to improve the switching speed of the VCSEL. The rise and fall times of the VCSEL driver current are required to be less than 1 ns. In order to minimize the power supply noise on the opto-board, the VDC should also have constant current consumption independent of whether the VCSEL is in the bright (on) or dim (off) state.

Figure 1(a) shows a block diagram of the VDC circuit. An LVDS receiver converts the differential input into a single-ended signal. The differential driver controls the current flow from the positive power supply into the anode of the VCSEL. An externally controlled current, $I_{set}$, determines the amplitude of the VCSEL current (bright minus dim current), while an externally controlled voltage, tunepad, determines the dim current. The differential driver contains a dummy driver circuit that in the VCSEL dim state draws an identical amount of current from the power supply as is flowing through the VCSEL in the bright state. This enables the VDC to have constant current consumption.

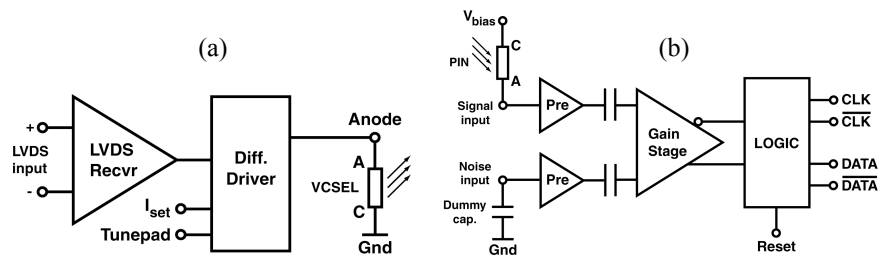

Figure 1. Block diagrams of the VDC (a) and DORIC (b) circuits.

## 3. DORIC Circuit

The DORIC decodes the BPM encoded clock and data (command) signal received by a PIN diode. The BPM signal is derived from the 40 MHz beam crossing clock by sending only transitions corresponding to clock leading edges. In the absence of data bits (logic level 1), this results simply in a 20 MHz signal. Any data bit in the data stream is encoded as an extra transition at the clock trailing edge.

The amplitude of the current from the PIN diode is expected to be in the range of 40 to 1000 µA. The 40 MHz clock recovered by the DORIC is required to have a duty cycle of (50 ± 4)% with a total timing error of less than 1 ns. The bit error rate of the DORIC circuit is required to be less than $10^{-11}$ at end of life.

Figure 1(b) shows a block diagram of the DORIC circuit. In order to keep the PIN bias voltage (up to 10 V) off the DORIC chip, we employ a single-ended preamp circuit to amplify the current produced by the PIN diode. Since single-ended preamp circuits are sensitive to power supply noise, we utilize two identical preamp channels: a signal channel and a noise cancellation channel. The signal channel receives and amplifies the input signal from the anode of the PIN diode, plus any noise picked up by the circuit. The noise cancellation channel amplifies noise similar to that picked up by the signal channel. This noise is then subtracted from the signal channel in the differential gain stage. To optimise the noise subtraction, the input load of the noise cancellation channel should be matched to the input load of the signal channel (PIN capacitance) via an external dummy capacitance.

## 4. Results From IBM 0.25 μm Submissions

The pixel detector design of the VDC and DORIC takes advantage of the development work for similar circuits [2] used by the outer detector, the SemiConductor Tracker (SCT). Both SCT chips attain radiation-tolerance by using bipolar integrated circuits (AMS 0.8 μm BICMOS) and running with high currents in the transistors at 4 V nominal supply voltage. These chips are therefore not applicable for the higher radiation dosage and lower power budget requirements of the pixel detector.

We originally implemented the VDC and DORIC circuits in radiation-hard DMILL 0.8 μm technology with a nominal supply voltage of 3.2 V. An irradiation study of the DMILL circuits in April 2001 with 24 GeV protons at CERN showed severe degradation of circuit performance. We therefore migrated the VDC and DORIC designs to standard deep submicron (0.25 μm) CMOS technology which has a nominal supply voltage of 2.5 V. Employing enclosed layout transistors and guard rings [3], this technology promises to be very radiation hard. We have had five prototype runs using 3-metal layouts over the course of two years with IBM as the foundry. For the engineering run, the layout was converted to 5-metal layouts in order to share the wafers with the MCC chips for cost saving.

We have extensively tested the chips to verify that they satisfy the design specifications. Figure 2 shows the VCSEL current generated by the VDC as a function of the external control current $I_{set}$. The saturation at high $I_{set}$ is due to the large serial resistance of the VCSEL. The dim current is close to the design value of 1 mA. The performance of the chips on opto-boards has been studied in detail. Each opto-board contains seven optical links. The typical PIN current thresholds for no bit errors are low, ~20-50 µA, and independent of the activity in adjacent DORIC channels. Channels with thresholds above 40 µA can be significantly reduced by adding a capacitor to the noise canceling channel. We have thus demonstrated the principle of the noise canceling circuit.

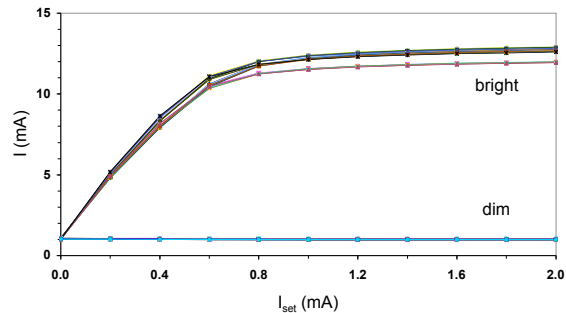

Figure 2. Measured VDC output current vs $I_{set}$ for eight four-channel VDC chips.

## 5. Irradiation Studies

In the last three years, we have performed three irradiations of the VDC and DORIC chips produced by the deep submicron process. We used 24 GeV protons at CERN for the study. No significant degradation in the performance of the chips was observed. In the following we describe the results from the last irradiation, August 2003, in more detail.

We used two setups during the irradiation. In the first setup, we performed electrical testing of four VDC and four DORIC chips. For the DORICs, we monitored the rise/fall times, clock duty cycle, and minimum input signal for no bit errors. For the VDC, we monitored the drive current for the bright and dim states. As expected, no significant degradation was observed up to 80 Mrad (2.6 x $10^{15}$ p/cm$^2$).

In the second setup, we tested the performance of four opto-boards. In the control room, we generated bi-phase mark encoded pseudo-random signals for transmission via 25 m of optical fibers to the opto-boards. The PIN diodes on the opto-boards converted the optical signals into electrical signals. The DORIC then decoded the electrical signals to extract the clock and command LVDS signals. The LVDS signals were fed into the VDCs and converted into signals that were appropriate to drive the VCSELs. The optical signals were then sent back to the control room for comparison with the generated signals. We remotely moved the opto-boards via a shuttle out of the beam to anneal the

VCSELs for ~19 hours after ~5 hours (~5 Mrad) of irradiation. The PIN current thresholds for no bit errors remained constant throughout the irradiation as shown in Fig. 3. For the optical power, we observed that the power decreased during the irradiation but partially recovered during the limited annealing period as expected. We expect the VCSEL power to almost completely recover after an extended annealing at the home institutions.

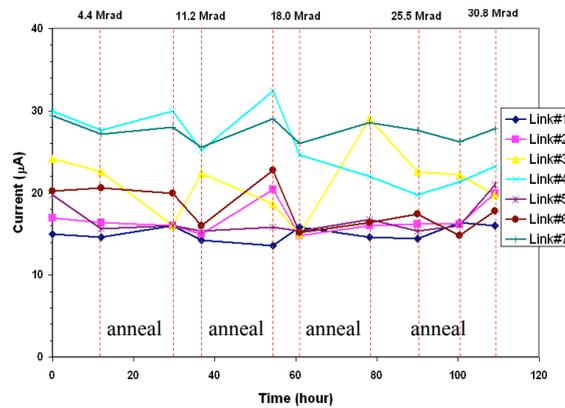

Figure 3: PIN current thresholds for no bit errors as a function of time for an opto-board.

## 6. Summary

We have developed VDC and DORIC chips in deep submicron (0.25 $\mu$m) technology using enclosed layout transistors and guard rings for improved radiation hardness. The circuits from the final design as well as the carrier, the opto-board, meet all the requirements of the optical readout system of the ATLAS pixel detector and further appear to be sufficiently radiation hard for ten years of operation at the LHC.

**Acknowledgements**

This work was supported in part by the U.S. Department of Energy under contract No. DE-FG-02-91ER-40690.